\documentstyle[12pt,psfig]{article}
\titlepage
 
\begin{document}
\setcounter{page}{1}
\title{New Scaling Law for Deuteron Production in Ultra-Relativistic
Nucleus Nucleus Collisions}
\author{H.\ Sorge \thanks{ E-mail: sorge@th.physik.uni-frankfurt.de }
\\ Institut f. theoretische Physik \\ J.\ W.\ Goethe Universit\"at
Frankfurt, Germany \\
[1em] J.L. Nagle, and B.S.\ Kumar \\ A.W.Wright
Nuclear Structure Laboratory, Yale University, \\ New Haven, CT, U.S.A.  }
\date{}
\maketitle
\begin{abstract}
  Deuteron production in S and Pb induced collisions at beam energies
  of 200 and 160 AGeV is studied in the framework of the transport
  theoretical approach RQMD.  Strong transverse flow invalidates the
  differential coalescence formula in momentum space.  The transverse
  momentum integrated $d$ yields scale in a broad rapidity interval
  with the squared proton densities and inversely with the produced
  particle rapidity densities.  This kind of scaling can be linked to
  constant relative sizes of nucleon and pion sources at freeze-out.
  With increasing projectile mass the nucleon source blows up stronger
  than the pion source. As a result, the scaled deuteron densities
  drop in central Pb+Pb collisions by 15 percent as compared to S
  induced reactions.
\end{abstract}
\hspace{24em}   UFTP-385/95

\section{Introduction}
High energy nucleus-nucleus collisions have been explored for several
years as a means of creating novel states of matter \cite{QM95,QM93}.
Observables which are directly related to the space-time structure of
the reactions are of particluar interest, because they allow for a
final `snapshot' of the hadron emitting source in Minkovski space.
The cluster production rate depends on the distances between nucleons
in the formation process.  Since only nucleons with similar positions
and momenta fit into the cluster wave function, the formation
probability for a cluster with mass $A$ is actually approximately
proportional to the
average of the
($A$-1)-power of the nucleon phasespace density.
 The cluster production rates do not change once
entropy production has ceased in the collision \cite{CK86}.
Hence, clusters are sensitive to the state of the system at
much earlier times than when they are formed.
In this paper, we suggest observables whose measurement will help
unravel the relative sizes of proton
and pion sources in $AA$ collisions
at energies above 100 AGeV.
The  final source sizes reflect a different (de-)coupling
strength of particle species in the created medium \cite{HAG94,CLE93}
and also differences in their production mechanisms \cite{SOR95}.
Just after hadronization
the baryon source at
midrapidity
might have a size  different from   the source of secondaries.
For instance,
 the transverse size of the baryon source is expected to be
 relatively smaller
in symmetric $AA$ collisions,
 because projectile stopping is a function of the target
thickness.  \footnote{
In the average a nucleon has to propagate through a lot of
nuclear material -- on the order of 12 fm at nuclear ground state
density -- to end up in the central rapidity region at CERN energy
(200AGeV).  The longitudinal velocity has to be degraded by three
units of rapidity.  A single inelastic $NN$ collision shifts a nucleon
only by about 0.7 units. The effective shift in $pA$ collisions is
even smaller, approximately 0.5 per collision \cite{BUS88}.  }
Since nucleons have a larger cross section than pions in
hadronic matter which contains only a small baryon fraction,
they decouple from the medium at lower density or -- equivalently --
show a larger size at freeze-out.  It would be interesting to know
the dominating
effects
which determine the final size of the
proton source as compared to the total source,
mostly pions at large beam energy.

\section{Scaling Laws}
It was suggested a long time ago that the invariant yields of light
nuclei with mass $A$ scale with the $A$-th power of the proton yields
\begin{equation}
E_A{d^{3}N_A \over{dp_A^3}} = B_A \left(E_p {d^3N_p \over{dp^3_p}}
\right)^A
\label{eq1}
\end{equation}
with a universal scale factor $B_A$ determined by the nuclear
parameters of the ingoing projectile and target \cite{BUT63}.  This
scaling relation was very successful in describing the yields of
light nuclei in nucleus-nucleus interactions at Bevalac and SIS
energies around 1 AGeV \cite{CK86}.  Note that the scaling factor
$B_A$
has been chosen originally as
volume-independent. This may indicate that
the expansion at low energies proceeds practically isentropically.
However, scaling is violated
by a factor of up to 40 in
nuclear collisions at the beam energy of 10-15 AGeV \cite{BAR93}.
Production
of additional particles which has been neglected in the scaling
relation eq.~\ref{eq1} may
impact cluster
production at the higher beam energy.  Since additional particles can
interact with the original nucleons as well, the system must expand
to a lower density before it can break up.  In particular, entropy
generating processes like the formation and decay of resonances play
an
important role in the space-time evolution of the heavy ion collisions
according to calculations with the transport model RQMD
\cite{SOR91,HOF94}.  Recently, the spectra of deuterons and protons
produced in Si(14.5AGeV) collisions with a heavy target which
 were measured
by the groups E814 \cite{BAR93} and E802 \cite{ABB94} have been
analysed and compared to calculations with RQMD \cite{NAG94}.  The
data were interpreted as evidence for considerable transverse
expansion beyond the volume given by the geometric overlap region of
the colliding nuclei.  The transverse expansion is most pronounced for
central collisions.  Of course, this is expected if the interactions
between original nucleons and secondary particles are of importance
for cluster formation.

In this paper we suggest a new scaling relation for deuteron yields
which may be appropriate for ultrarelativistic energies.  We suggest
that the $d$ yields scale inversely with the produced particle
rapidity density in $AA$ collisions at sufficiently large beam
energies.  We express the parameter $B_d$ as
\begin{equation}
 \label{eq2}
   B_d = c \cdot (dN^{-}/dy)^{-1}
  \quad  ,
\end{equation}
 which may reflect better the conditions of nuclear interactions at
ultrarelativistic energies than $B_A$ definitions ignoring particle
production.
The so defined $B_d$ contains implicitly a dependence on
the source size of the reaction, because the particle rapidity density
\footnote{ We have chosen -- somewhat arbitrarily -- the negative
particles for the scaling law.  Note that mesons are populating the
three charge states in approximately equal amounts at
ultrarelativistic energies.  We have included feed-down from weak
decays (except $\Lambda $ and $K_S$) in the negatives, because it is
experimentally much easier to subtract hyperon feed-down from measured
proton yields than to control the amount of meson decays, e.g.\ of
$\eta$ and $\eta '$.  Our conclusions in this paper are not affected
by this procedure.  } is proportional to the freeze-out volume, if
hadrons freeze-out at constant density.  The radius parameters of
Bose-Einstein correlations scale rather well with the rapidity density
as $\sim (dN^{+/-}/dy)^{1/3}$.
This dependence was reported for high energy $hh$ and $AA$ reactions
by various experimental groups (AFS, UA1, E802, NA35 and NA44) and is
consistent with freeze-out at constant density.
 (See ref. \cite{STO91} for a review on recent HBT results.)
 One could infer from
a $B_d$ dependence on the produced particle rapidity density as in
eq.~\ref{eq2} that the nucleon source size scales with the produced
particle source.  By using eq.~\ref{eq2} we hope to incorporate the
most important aspect of particle production on nuclear coalescence
probabilities.  The formula eq.~\ref{eq2} should be applied only for
collisions which are dominated by particle production, i.e.\ with a
ratio $N_\pi$$/$$N_B$ much larger than 1.  Therefore we test the
scaling relation for A on A reactions which are currently under
experimental study at the CERN-SPS (160-200AGeV).  Here we are using
phasespace distributions calculated with the RQMD model (version 1.09)
\cite{SOR95,SOR95ZFP,BER94} in combination with the Wigner function
method for cluster coalescence \cite{REM81,GYU83,MAT95}.
There are only prelimimary experimental data  on cluster and
anti-cluster production in $AA$ collisions at these energies available
from the NA44 and NA52 groups
\cite{NA4495,NA5295}. Comparisons with RQMD
calculations are under way \cite{NAG95APS,GIL95}.

RQMD is a multiple-collision approach
in which the secondary particles emerge as fragmentation products of
decaying color strings, ropes and resonances. Afterwards they may
interact with each other and the ingoing nucleons.  Nuclear cluster
formation is not included dynamically in RQMD. It is added after the
strong interactions have ceased and the particles are streaming
freely.  The deuteron yields are calculated from the product of
proton and neutron source functions
($g_p$ and $g_n$) at freeze-out
and the
coalescence factor, integrated over phasespace:
\begin{equation}
\label{Nd} N_d = \int d^4 x_1 \, d^4 x_2 \int d p_1 \, d p_2 \:
g_p(x_1,\vec{p}_1) g_n(x_2,\vec{p}_2) p_d(1,2) \quad .
\end{equation}
The coalescence factor $p_d$ is taken as
\begin{equation} \label{pd} p_d= 3/8 \cdot
f_W^d(\vec{x}_{CMS},\vec{p}_{CMS}) \quad .
\end{equation}
$f_W^d$ denotes the Wigner transform of the deuteron wave function,
with center-of-mass motion
removed. Its arguments are the distance between the nucleons at the
larger of the two freeze-out times and the relative momentum, all
values evaluated in the deuteron
center of mass frame
where $\vec{p}_1 +\vec{p}_2 =\vec{0}$.
Here we use a harmonic oscillator wave function with parameters as
employed for $d$ coalescence calculations in \cite{GYU83}.  The
prefactor in eq.~\ref{pd} comes from statistical isospin and spin
projection of a $np$ pair onto the corresponding deuteron quantum
numbers.  In the semiclassical RQMD model the final particle energies
are set on-shell.  Thus it is implicitly assumed by applying
eq.~\ref{Nd} that the true quantum-mechanical source functions at
freeze-out have negligible energy dependence on the scale of the
deuteron binding energy around the mass shell.

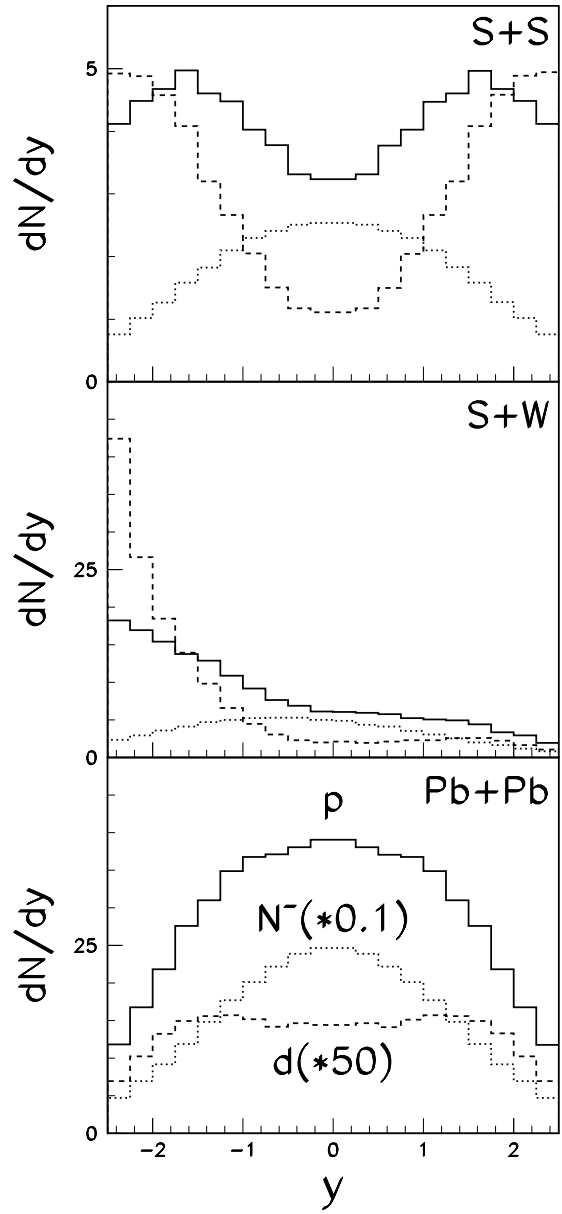
 \begin{figure}[h]
 \vspace{2.0cm}

 \caption
 [
  ]
 {
  \label{daadndy}
 }
 \end{figure}

The calculated rapidity distributions of deuterons and protons
(without feed-down from weak decays) are displayed in
fig.~\ref{daadndy}.  The RQMD calculations have been carried out for
three different systems, S on S (impact parameter b$<$1 fm), S on W
(b$< $4 fm), both at a projectile energy of 200 AGeV, and Pb on Pb
(b$< $1 fm), with beam energy 160 AGeV.  The different shapes of the
baryon distributions
reflect the dependence of the stopping
power on target and projectile mass number and are discussed in
\cite{SOR95ZFP}.  The calculated $d$/$p$ ratios are less than one
percent in the central rapidity region, clearly smaller (by an order
of magnitude) than the data measured at the lower beam energies of
10-15 AGeV \cite{ABB94}.

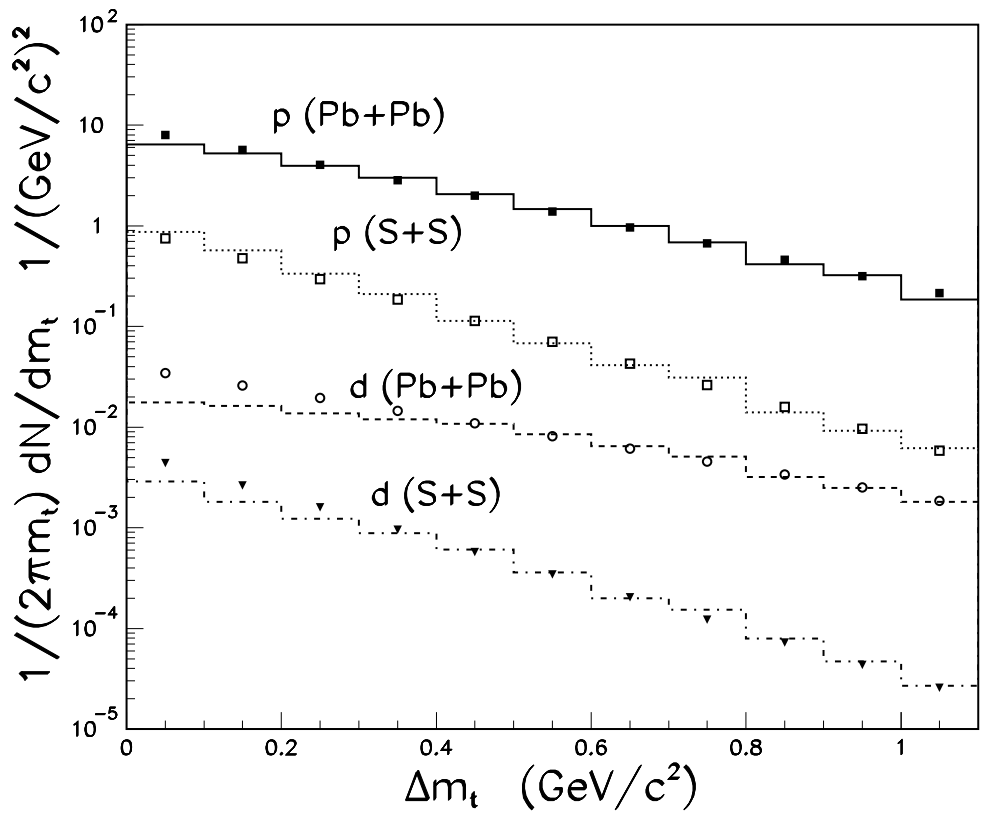
 \begin{figure}[h]
 \vspace{2.0cm}

 \caption
 [
  ]
 {
  \label{daadndmt}
 }
 \end{figure}

A main motivation to compare calculated deuteron yields to scaling
relations like eqs.~\ref{eq1},\ref{eq2} is that some basic aspect of
the dynamics -- more particles need a larger freeze-out volume -- is
accounted for.  Other collective dynamical effects in
ultrarelativistic collisions like collective flow may lead to scaling
violations.  Indeed, an immediate consequence of eq.~\ref{eq1} is that
the transverse mass spectrum of the nucleon clusters will have the
same slope as the primordial nucleon spectra.  This results from
eq.~\ref{Nd} only, if
the freeze-out distribution
$g_N$ factorizes into a product of separate
functions of position and momentum, which precludes flow.  Transverse
mass spectra calculated from RQMD are shown in
fig.~\ref{daadndmt} (histograms).
They are compared to fits with a Boltzmann distribution
(symbols). The primordial proton spectra are  well
described by the thermal distributions with `apparent temperature'
(slope parameter)
180 MeV for S+S and 232 MeV for Pb+Pb collisions.
At large $m_t$ values
the
deuteron  distribution
is fitted by the Boltzmann distribution
with  same slope parameter as for the protons
only in  the S+S case.
The slope parameter which is needed for the
tail of the  deuteron $m_t$ distribution
 from  Pb+Pb collisions has  quite a different value,
 300 MeV.
Furthermore, the calculated deuteron distributions
differ markedly from a thermal shape if the spectra at low  $m_t$ values
are taken into account. They exhibit
a  convex curvature (shoulder-arm shape)
which is characteristic for the presence
 of transverse flow.
We may therefore
conclude that the coalescence formula eq.~\ref{eq1} is in
contradiction with the calculated momentum distributions and therefore
useless for further consideration here.

However, we can consider the validity of a coalescence formula similar
to eq.~\ref{eq1} in a
modified form, for the particle yields after
integration over transverse momentum
\begin{equation}
\label{coala}
dN_d/dy=B_d \cdot (dN_p/dy)^2 \quad .
\end{equation}
A $B_d$
dependence which is inverse to the volume $V$=$ \int d^3 \sigma _\mu
u^\mu (x)$
of the baryon source
after freeze-out can be derived from colescence formula
eq.~\ref{Nd}
even in the presence of flow. The velocity
field $\vec{u}(x)$ acquires nonzero components in this case.
(The 4-vector $ d^3 \sigma _\mu $ has only
  a nonzero value for the 0-th  component
  equal to $d^3x$ in the local rest system defined by
 $\vec{u}' (x)=\vec{0}$.)
The derivation  holds under fairly
general conditions: not too large spreading of emission times, local
statistical equilibrium, constant local baryon density and freeze-out
temperature, and small flow gradients on the scale of the deuteron
size.  The last assumption is needed for a factorization of $g_N$ to
be valid locally.
Assuming local statistical equilibrium
at some freeze-out temperature the pion yield is proportional
to its source volume as well.
It is straightforward under these assumptions to calculate the relation
between coalescence parameter $B_d$ and produced particle
yields
for a hadron  gas in equilibrium.
 We will not do this here, but instead we will discuss to which extent
scaling holds according to the RQMD results.

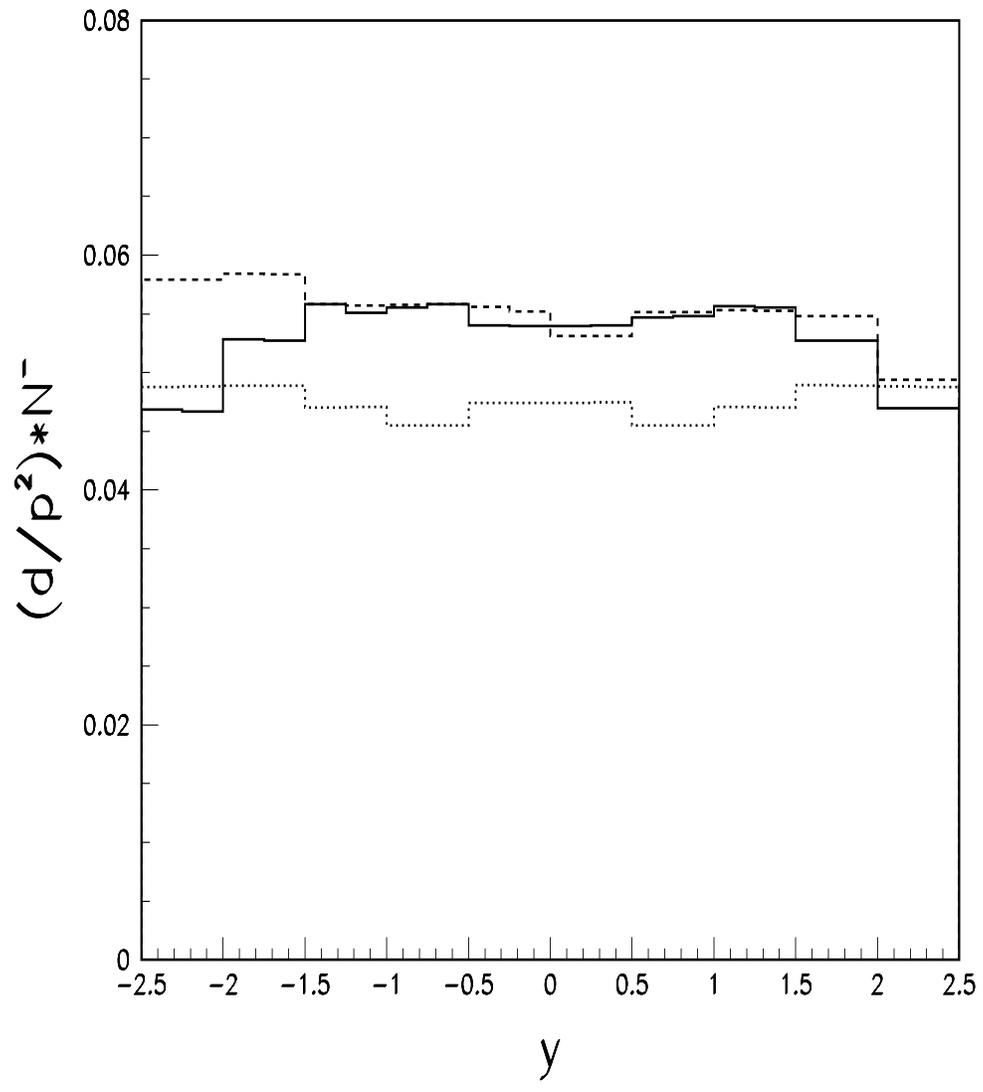
 \begin{figure}[h]
 \vspace{2.0cm}

 \caption
 [
  ]
 {
  \label{dscale}
 }
 \end{figure}

The scaled $d$ rapidity density $[dN_d/dy/(dN_p/dy)^2]\cdot dN^-/dy$
will be constant if eqs.~\ref{eq2},\ref{coala} are satisfied.  This
ratio -- calculated from the RQMD results -- is displayed in
fig.~\ref{dscale} as a function of rapidity for the three colliding
systems under consideration.  Indeed, the scaled $d$ rapidity density
turns out to be remarkably independent of rapidity in a broad interval
of about three units.  (It is quite natural that scaling breaks down
in the projectile and target fragmentation region, because the basic
assumptions, in particular meson dominance, are not fulfilled here.)
There is essentially no target dependence in S induced reactions by
going from a S to a W target.
While
collisions with  central impact parameters ($b$$<$1 fm)
were taken  for the symmetric systems,
the S+W events were generated in an impact parameter range
up to 4 fm. We have claculated
the $d$ rapidity densities for S+W in
 different event classes defined by  centrality
($b$$<1$ fm, 1 fm$<$$b$$<2$ fm, and so on).
We find that the scaled $d$ distributions do not depend on
$b$ within this impact parameter range.
While scaling holds separately for the
Pb+Pb collisions, it becomes clear from fig.~\ref{dscale} that the
scaled $d$ yields are smaller in central Pb+Pb collisions -- by about
15 \% -- than in the S+A reactions. This points towards a relative
increase of the nucleon source in comparison to the produced particle
source.

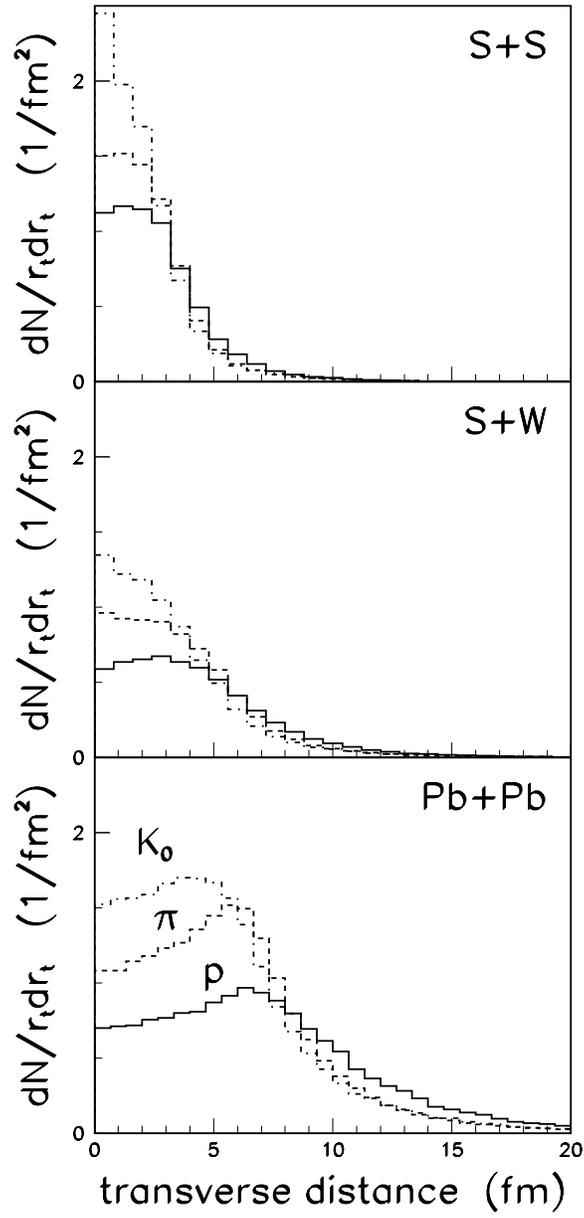
 \begin{figure}[h]
 \vspace{2.0cm}

 \caption
 [
  ]
 {
  \label{trdaa}
 }
 \end{figure}

The relative increase of the nucleon source in central Pb+Pb
collisions can be checked directly by looking at the RQMD freeze-out
distributions (see fig.~\ref{trdaa}).  Fig.~\ref{trdaa} shows
transverse freeze-out distributions for pions, protons and kaons at
midrapidity
($y_{mid}$-1$<$$y$$<$$y_{mid}$+1).
The
relative size of the proton source grows relatively to the meson
source by going from S+S to Pb+Pb.  While the ratio of proton to pion
transverse RMS radius (at midrapidity) is 1.05 in S+S reactions
(5.4 fm:5.1 fm), it increases to 1.19 in central Pb+Pb collisions
(10.3 fm:8.6 fm).  The system has expanded considerably before
freeze-out, because the initial transverse RMS sizes of S (Pb) have
values 2.4 (4.5) fm only.  RQMD shows no indications for a narrower
transverse source size of midrapidity protons, just the opposite.  The
RQMD freeze-out distributions are not consistent with scaling in any
power of number of participants, produced particle or nucleon rapidity
density.  While the source distributions have approximately a Gaussian
shape in S+S (and S+W), the transverse flow is clearly visible in the
distributions for Pb+Pb.  A blast wave has transported much of the
material into the outer space leaving a hole in the interior region.
Inspection of the reaction history in RQMD reveals why this effect
depends -- to some extent -- on the absolute size of the reaction
region.  Starting at
a value of zero, the transverse baryon flow develops
initially more slowly than the pion flow.  The responsible factor here
is the mass difference between baryons and mesons which is not
compensated by sufficiently strong interaction to form really one
fluid.  The time averaged flow of particles freezing-out is affected
by the same mechanism.  The different strength of the transverse flow
is weighted more heavily in smaller systems even if the particle
density was initially the same, because the meson source has
evaporated and the interactions with baryons have ceased earlier than
in central collisions of heavy ions like Pb+Pb.

\section{Conclusions}

We have studied deuteron production based on
RQMD calculations for S and Pb induced collisions at beam energies of
200 and 160 AGeV.  We find violation of the usual differential
coalescence formula in momentum space, in particular in central Pb on
Pb collisions, which is caused by the presence of strong transverse
flow.  Transverse momentum integrated $d$ yields scale in a broad
rapidity interval with the squared proton densities and inversely with
the produced particle rapidity densities.  We have argued that this
kind of scaling can be linked to constant relative sizes of nucleon
and pion source at freeze-out.  With increasing the projectile mass,
we find a blow-up of the nucleon source which is stronger than for the
pion source.  Therefore the scaled deuteron densities drop in central
Pb+Pb collisions by 15 percent as compared to S induced reactions.
The calculations which have been presented here demonstrate the
usefulness of cluster measurements to extract information on the
created source -- particularly its size and the strength of flow -- in
ultrarelativistic nucleus-nucleus collisions.

\section{ Acknowledgements }

H.\ S.\ is grateful to M.\ Leltchouk for discussions about collective
flow. He also thanks
J.\ Simon-Gillo and R.\ Mattiello
 for stimulating discussions.
This work was supported in part by GSI and DFG, and the
U.S. DoE, grant number DE-FG02-91ER-40609.

\newpage

{\noindent  \LARGE   Figure Captions:}
\vspace{1.0cm}

{\noindent \large Figure 1: }

{\noindent
Final rapidity distributions
of negatively charged hadrons, primordial protons and deuterons
calculated from RQMD (version 1.09).
The RQMD calculations have been carried out for
three different systems, S on S (impact parameter b$<$1 fm), S on W
(b$< $4 fm), both at a projectile energy of 200 AGeV, and Pb on Pb
(b$< $1 fm), with beam energy 160 AGeV.
 The  distributions are displayed as histograms,
 straight line for protons, dashed for deuterons
 (multiplied with 50) and dotted for negatives
 (multiplied with 0.1).
 The rapidity is calculated in the equal-speed-system
 of projectile and target.
  The generated events for the symmetric systems have been reflected
  $z \rightarrow -z$  in order to improve statistics.
}
\vspace{0.5cm}

{\noindent \large Figure 2: }

{\noindent
  Transverse mass spectra
of  primordial protons and deuterons
  in the rapidity
     window $y_{mid}-0.5<y<y_{mid}+0.5$.
  The  spectra have been calculated for S+S and Pb+Pb
   using the RQMD model
   under the same conditions as explained in
   caption to fig.~1.
 They are represented as  histograms,
  protons from Pb+Pb (straight line),
  deuterons from Pb+Pb (dashed line),
  protons from S+S (dotted line),
  and deuterons from S+S (dashed-dotted line).
  The calculated spectra are compared to
  Boltzmann distributions
  $\sim m_t \cdot \exp (-m_t/T) $ with slope
  parameters fitted to the large $m_t$ tail
  of the distributions, with $T$ parameters
  (from top to bottom) 232, 180, 300 and 180 MeV.
}
\vspace{0.5cm}

{\noindent \large Figure 3: }

{\noindent
The scaled $d$ rapidity density $[dN_d/dy/(dN_p/dy)^2]\cdot dN^-/dy$
as a function of rapidity.
The RQMD calculations have been carried out for the same
three  systems as in fig.~1, S on S
(straight line), S on W (dashed histogram)
and Pb on Pb (dotted histogram).
}
\vspace{0.5cm}

{\noindent \large Figure 4: }

{\noindent
 Distribution of
 transverse
 distances (to the collision center)
 $1/r_t dN/dr_t$
 at freeze-out
 for different particle species in
 S on S (top), S on W (middle) and Pb on Pb (bottom)
 collisions.
 Only particles around central rapidity ($y_{mid}$$\pm$1) are
 taken into account, primordial
  protons (straight line),
  pions (dashed histogram),
  and neutral (anti-)kaons $K^0$+$\bar{K^0}$
  (dotted histogram).
  Note that the pion and kaon distributions
  have been renormalized that the integral gives
  the same yield as for the protons.
}
\vspace{0.5cm}

 \newpage

\end{document}